\documentclass[british]{INTERSPEECH2023_arxiv}

\usepackage[utf8]{inputenc}
\usepackage[T1]{fontenc}
\usepackage[british]{babel}

\usepackage{pifont}
\newcommand{\cmark}{\ding{51}}

\usepackage{graphicx}
\usepackage{epsfig}
\usepackage{adjustbox} 
\usepackage{xcolor}
\graphicspath{{./figures/}}
\DeclareGraphicsExtensions{.pdf,.png,.jpeg,.jpg}

\usepackage{amsmath}
\usepackage{amssymb}
\usepackage{bm} 
\usepackage{esint} 
\usepackage{commath} 
\usepackage{nicefrac}

\usepackage{subcaption} 

\usepackage{booktabs} 
\usepackage{array} 
\usepackage{makecell}

\usepackage{multirow} 
\usepackage{rotating} 

\usepackage{enumitem} 
\setlist{nolistsep} 

\usepackage[unicode=true,pdfusetitle,%
  pdfauthor={Shivam Mehta, Ambika Kirkland, Harm Lameris, Jonas Beskow, Éva Székely, Gustav Eje Henter},%
  pdfkeywords={probabilistic TTS, acoustic modelling, hidden Markov models, Glow, invertible post-net},%
  bookmarks=true,bookmarksnumbered=true,bookmarksopen=false,%
  breaklinks=true,pdfborder={0 0 0},backref=false,colorlinks=true,citecolor=blue]{hyperref}

\usepackage[capitalize]{cleveref} 
\Crefname{section}{Section}{Sections}
\crefname{section}{Sec.}{Secs.}
\Crefname{table}{Table}{Tables}
\crefname{table}{Tab.}{Tabs.}

\newcommand{\bb}[1]{\boldsymbol{#1}}

\renewcommand{\mid}{\,\ifnum\currentgrouptype=16 \middle\fi|\,}

\newcommand{\real}{\mathbb{R}}

\newcommand{\webpageurl}{https://shivammehta25.github.io/OverFlow/}
\newcommand{\webpageurltext}{shivammehta25.github.io/OverFlow/}

\newcommand{\yes}{\cmark}

\newcommand{\tablebf}[1]{%
\pdfliteral direct {2 Tr 0.5 w}
#1%
\pdfliteral direct {0 Tr 0 w}%
}

\let\oldmarginpar\marginpar
\renewcommand\marginpar[1]{\-\oldmarginpar[\raggedleft\footnotesize #1]%
{\raggedright\footnotesize #1}}

\interspeechcameraready

\title{OverFlow: Putting flows on top of neural transducers for better TTS}

\name{Shivam Mehta, Ambika Kirkland, Harm Lameris, Jonas Beskow, {\'E}va Sz{\'e}kely, Gustav Eje Henter}
\address{Division of Speech, Music and Hearing, KTH Royal Institute of Technology, Stockholm, Sweden}
\email{\{\href{mailto:smehta@kth.se}{smehta}, \href{mailto:kirkland@kth.se}{kirkland}, \href{mailto:lameris@kth.se}{lameris}, \href{mailto:beskow@kth.se}{beskow}, \href{mailto:szekely@kth.se}{szekely}, \href{mailto:ghe@kth.se}{ghe}\}\href{mailto:ghe@kth.se}{@kth.se}}

\begin{document}
\pagestyle{plain}
\maketitle
\begin{abstract}
Neural HMMs are a type of neural transducer recently proposed for sequence-to-sequence modelling in text-to-speech. They combine the best features of classic statistical speech synthesis and modern neural TTS, requiring less data and fewer training updates, and are less prone to gibberish output caused by neural attention failures. In this paper, we combine neural HMM TTS with normalising flows for describing the highly non-Gaussian distribution of speech acoustics. The result is a powerful, fully probabilistic model of durations and acoustics that can be trained using exact maximum likelihood. Experiments show that a system based on our proposal needs fewer updates than comparable methods to produce accurate pronunciations and a subjective speech quality close to natural speech.
\end{abstract}
\section{Introduction}
\label{sec:intro}
It was recently demonstrated \cite{mehta2022neural} that sequence-to-sequence neural text-to-speech (TTS) can be improved by replacing conventional neural attention with a so-called neural hidden Markov model, or \emph{neural HMM} \cite{tran2016unsupervised}, which is a kind of \emph{neural transducer} \cite{yu2016online}.
The resulting model class combines the best features of classic, HMM-based TTS and modern neural sequence-to-sequence models \cite{watts2019where,mehta2022neural}:
models are fully probabilistic and can be trained to maximise the exact sequence likelihood, and the use of left-to-right no-skip HMMs ensures each phone in the input is spoken in the correct order, a property known as \emph{monotonicity}.
This directly addresses issues due to non-monotonic attention \cite{he2019robust}, which make many attention-based neural TTS models prone to speaking random gibberish, and generally require a lot of data and updates to learn to speak properly.
The sequential nature of autoregressive synthesis is advantageous on devices with limited parallelism, and also lends itself well to incremental and streaming applications of TTS, as shown in \cite{chen2021speech}.

However, TTS based on the neural HMM framework has yet to live up to its full potential.
In particular, the systems demonstrated in \cite{yasuda2019initial,chen2021speech,mehta2022neural} include restrictive assumptions that state-conditional emission distributions are Gaussian (equivalent to an L2 loss) or Laplace (equivalent to an L1 loss).
This makes for weak models of human speech, since natural speech signals follow a highly complex probability distribution.

We present OverFlow, which adds normalising flows on top of neural HMM TTS 
to better describe the non-Gaussian distribution of speech parameter trajectories.
This gives a fully probabilistic
model of acoustics and durations that, unlike most flow-based acoustic models \cite{miao2020flow,kim2020glow,shih2021rad,miao2021efficienttts}, uses autoregression to enable long-range memory.
Experiments show that the model quickly learns to produce accurate pronunciation and good synthesis quality, outperforming comparable systems based on related
approaches like Tacotron 2 \cite{shen2018natural} and Glow-TTS \cite{kim2020glow}.
For audio and code see \href{\webpageurl}{\webpageurl}.


\begin{table}[t!]
\centering
\begin{tabular}{@{}lccccccc@{}}
\toprule 
\multicolumn{1}{r}{Approach:} & \cite{shen2018natural,wang2017tacotron,li2019neural} & \cite{kim2020glow} & \cite{miao2021efficienttts} & \cite{valle2021flowtron} & \cite{chen2021speech} & \cite{mehta2022neural,yasuda2019initial} & Prop.\tabularnewline
\midrule
Autoregression & \yes &  &  & \yes & \yes & \yes & \yes\tabularnewline
Post-net & \yes & N/A & N/A &  &  &  & \yes\tabularnewline
Monotonic &  & \yes & \yes &  & \yes & \yes & \yes\tabularnewline
Attention-free &  & \yes &  &  & \yes & \yes & \yes\tabularnewline
Fully prob. &  &  &  & \yes &  & \yes & \yes\tabularnewline
Flows &  & \yes & \yes & \yes &  &  & \yes\tabularnewline
\bottomrule
\end{tabular}
\caption{Overview of prior acoustic models and the proposed method. ``Fully probabilistic'' is defined in Sec.\ \ref{fully_prob_def}. ``Attention-free'' refers to the method for aligning input symbols to output frames, and does not consider, e.g., Transformer self-attention.
}
\label{tab:overview}
\vspace{-1.5\baselineskip}
\end{table}

\section{Prior work}
\label{sec:background}

\subsection{TTS with transducers and neural HMMs}
Neural HMMs \cite{tran2016unsupervised} are a kind of autoregressive HMM \cite{rabiner1989tutorial,
shannon2013autoregressive} where the state-conditional emission distribution and transition probability are both defined by a neural net.
This makes them vastly more powerful than classical HMMs.
The simplest versions of these models represent left-to-right no-skip HMMs, and are therefore monotonic.
As these models still satisfy classic hidden Markov assumptions, we can use the forward or Viterbi algorithms \cite{rabiner1989tutorial} to efficiently compute (or lower-bound) the log-likelihood, and then use stochastic gradient ascent on top of automatic differentiation to optimise it.
This makes these methods a compelling choice for autoregressive TTS \cite{mehta2022neural}.

Mathematically, neural HMMs are a kind of neural transducer \cite{yu2016online}.
Neural transducers are known for their strong automatic speech-recognition (ASR) performance in recent years (see, e.g., \cite{zhang2020pushing}), and were first used for TTS in SSNT-TTS \cite{yasuda2019initial}.
More recently, they were shown to handle both ASR and streaming TTS within a single model \cite{chen2021speech}.
Our approach differs from \cite{chen2021speech} in that it is probabilistic (unlike the modified training algorithm in \cite{chen2021speech}) and from \cite{yasuda2019initial} in that integrates quantile-based duration generation \cite{ronanki2016median,henter2017nonparametric} for more consistent speech-sound durations.
Most importantly, and unlike \cite{yasuda2019initial,chen2021speech,mehta2022neural}, we integrate normalising flows into this category of models in order to obtain a powerful probabilistic framework capable of describing the behaviour of speech acoustics.

\subsection{Normalising flows in TTS acoustic modelling}
\emph{Normalising flows} (a.k.a.\ \emph{flows}) \cite{papamakarios2021normalizing,kobyzev2021normalizing}
use deep learning to define highly flexible parametric families of probability distributions.
The idea is to create a complex probability distribution from a simple Gaussian by subjecting it to a series of nonlinear, invertible transformations based on neural networks.
Invertibility means that the change-of-variables formula can be used to compute and maximise the likelihood of the resulting model.

Flows have seen many applications across speech technology, but the first TTS acoustic models to integrate flows were Flowtron \cite{valle2021flowtron} and Flow-TTS \cite{miao2020flow}.
These were followed by Glow-TTS \cite{kim2020glow}, RAD-TTS \cite{shih2021rad}, and EfficientTTS \cite{miao2021efficienttts}.
Flowtron is 
an extension of Tacotron 2 with multiple speakers and global style tokens \cite{wang2018style} that puts a normalising flow inside the autoregressive decoder that generates the next output frame.
Like Tacotron 2, it uses neural attention to learn to speak and align.
Flow-TTS is a non-autoregressive architecture reliant on external alignments for training.
Glow-TTS improved on Flow-TTS by learning to speak and align at the same time, eliminating the need for external alignment tools.
It also introduced a dynamic-programming procedure based on the Viterbi algorithm from HMMs to enforce monotonic alignments during training.
RAD-TTS \cite{shih2021rad} integrated a separate normalising flow for duration modelling along with a more elaborate alignment mechanism, but did not attain the same speech quality as Glow-TTS.
EfficientTTS \cite{miao2021efficienttts}, finally, described two efficient mechanisms to encourage or enforce monotonic alignments with conventional dot-product attention, 
resulting in improved subjective scores.
\label{fully_prob_def}
Of these, only Flowtron is \emph{fully probabilistic}, as others lack a discrete-valued probabilistic model of sequence durations, and thus do not define a valid probability distribution over the set of all possible discrete-time output sequences.

Except Flowtron \cite{valle2021flowtron}, all the above flow-based TTS acoustic models are non-autoregressive (i.e., parallel).
This is advantageous on GPU servers.
However, autoregressive architectures have often had an edge over non-autoregressive ones in terms of probabilistic-modelling accuracy, e.g., in bits-per-pixel metrics on image benchmarks \cite{kingma2021variational}.
A combination of flows and non-autoregressive HMMs recently advanced the state of the art in phone recognition \cite{ghosh2020robust,ghosh2021normalizing}.
These findings motivate our innovation to combine flows with autoregression through neural HMMs, since both allow exact maximum-likelihood training and have demonstrated strong probabilistic-modelling results.

Flowtron \cite{valle2021flowtron} 
uses conventional neural attention that does not enforce monotonicity, making it prone to gibberish and difficult to train.
Unfortunately, the model cannot learn to speak by training on LJ Speech alone \cite{valle2021flowtron}, which would be needed for a fair comparison.
Glow-TTS \cite{kim2020glow} 
enforces monotonic alignments with an algorithm derived from HMMs, like our model.
We adopt their decoder architecture in our experiments and also compare to a Glow-TTS system.
Their method differs from our proposal in the lack of autoregression, in the use of Viterbi recursion (single path) over the complete forward algorithm (all paths), and in the use of a non-probabilistic duration model. 

Flows have also been used in end-to-end TTS, most prominently VITS \cite{kim2021vits}, which adds a flow-based duration model and a neural vocoder to Glow-TTS and trains on a compound loss.
This is powerful but, as our results show, much slower to train.

\subsection{Flows as an invertible post-net}
Many autoregressive TTS systems with regression losses, such as \cite{wang2017tacotron,shen2018natural,li2019neural},
use a greedy, sequential output generation procedure during synthesis.
To improve output quality, a so-called \emph{post-net} with non-causal
CNNs is then applied to ``go back'' over the sequence generated by the autoregressive model to enhance it.
Unfortunately, that standard post-net architecture is incompatible with
maximum-likelihood training of neural HMMs, and the absence of such a post-net negatively impacts speech quality \cite{mehta2022neural}.
By instead passing model output through an \emph{invertible} post-net (thus turning the model into a normalising flow), we are able to incorporate a post-net whilst still training the entire model to maximise exact sequence likelihood.
Our experiments 
use the decoder from Glow-TTS, which
leverages an architecture based on non-causal CNNs, evoking similarities to the post-nets in attention-based neural TTS \cite{wang2017tacotron,shen2018natural,li2019neural}.

As an alternative to a post-net, one may instead fine-tune the vocoder to produce more natural waveforms from unenhanced acoustic features.
However, this requires an additional training stage, and neural vocoder training is generally slow.

\begin{figure*}
\centering
\includegraphics[width=0.8\textwidth]{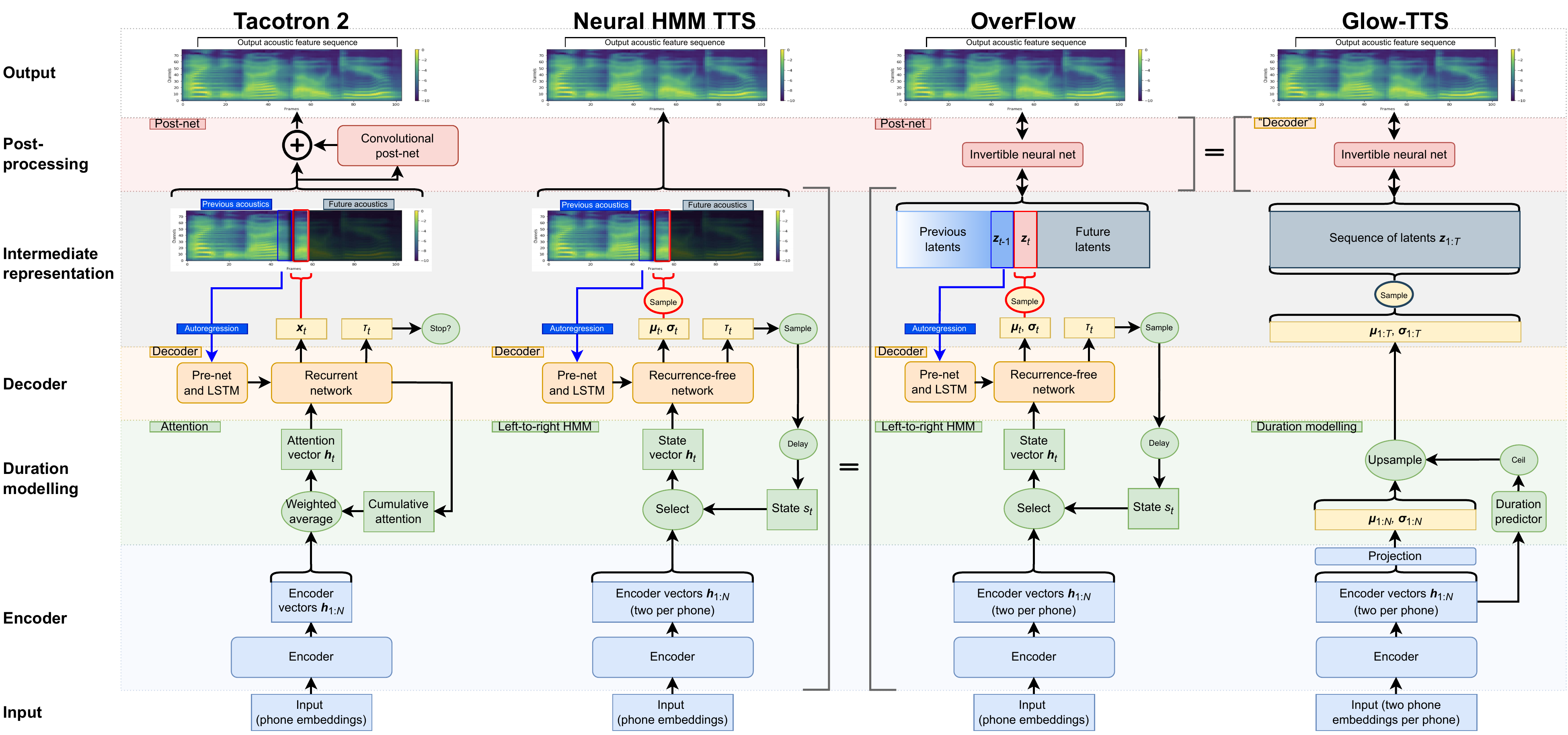}
\caption{Comparison of four systems from the experiments. Square brackets show how the OverFlow system passes the output of a neural HMM though the invertible neural network from Glow-TTS, similar to the post-net in Tacotron 2. Rectangular boxes are vectors and scalars, rounded boxes are learnt transformations (neural networks), and ovals are fixed mathematical operations.}
\label{fig:overview}
\vspace{-1.5\baselineskip}
\end{figure*}
\section{Method}
\label{sec:method}
Our proposed approach, OverFlow, 
is shown in Fig.\ \ref{fig:overview}.
In a nutshell, we apply an invertible neural network to the neural HMM output, 
greatly increasing the set of probability distributions that can be represented by the model.
Experiments in Sec.\ \ref{sec:experiments} 
confirm this significantly improves the output speech.
The rest of this section describes the two methods -- neural HMM TTS and normalising flows -- that make up the proposed approach.

Neural HMMs are a framework of probabilistic encoder-decoder sequence-to-sequence models that provide an alternative to conventional neural attention for aligning input values with output observations.
When used for TTS acoustic modelling as in \cite{yasuda2019initial,mehta2022neural}, the neural HMM \emph{encoder} converts the input vectors (e.g., phone embeddings) into a sequence of vectors $\bb{h}_{1:N}$ that define $N$ states in a left-to-right no-skip HMM.
Each input symbol is translated to a fixed number of vectors, for example two states per symbol.
The neural HMM \emph{decoder} network takes two inputs and returns two outputs, which we write as $(\bb{\theta}_t,\,\tau_t)=\mathrm{Dec}(\bb{h}_n,\,\bb{x}_{1:t-1})$.
The inputs are a state-defining vector $\bb{h}_{s_t}$ from the encoder, corresponding to the HMM state $s_t\in\{1,\,\ldots,\,N\}$ at timestep $t$, as well as the previously generated output acoustic frames $\bb{x}_{1:t-1}$.
The latter input makes the model autoregressive.
In practice, it is common to only provide the previous frame $\bb{x}_{t-1}$ 
as explicit input, and then process this input using a so-called \emph{pre-net}, along with an LSTM that ensures outputs from before $t-1$ also can influence the next frame.

The two decoder outputs are $\bb{\theta}_t$ (the parameters of the emission distribution) and a \emph{transition probability} $\tau_i\in[0,\,1]$.
The parameters $\bb{\theta}_t$ define a probability distribution for the next output frame $\bb{x}_t$.
Most neural HMMs assume that $\bb{x}_t$ follows a multivariate Gaussian distribution with diagonal covariance matrix.
In that case, the parameters $\bb{\theta}_t=(\bb{\mu}_t,\,\bb{\sigma}_t)$ are two vectors corresponding to the element-wise means and standard deviations of $\bb{x}_t$.
Meanwhile, $\tau_t$ defines the probability that the HMM transitions to the next state, so that $s_{t+1}=s_t+1$; else $s_{t+1}=s_t$.
Synthesis begins with $s_1=1$ and terminates when $s_t=N+1$.
To satisfy the Markov assumption over the hidden states $s_{1:T}$ -- and thus be a valid neural HMM -- the decoder output may not depend on the input vectors $\bb{h}_{1:t-1}$ at previous timesteps, which means that 
models can be trained to maximise the log-likelihood of training data, computed using the standard forward or Viterbi algorithms from classical HMMs \cite{rabiner1989tutorial}.
For a graphical overview of neural HMM TTS, see Fig.\ \ref{fig:overview}.

Neural HMMs describe a distribution over discrete-time sequences, typically 
assuming a Gaussian distribution for each frame.
Normalising flows \cite{papamakarios2021normalizing,kobyzev2021normalizing},
provide a method for turning simple \emph{latent} or \emph{source} distributions $\bb{Z}$ (e.g., Gaussians) into much more flexible \emph{target} distributions $\bb{X}$.
The idea is to apply an invertible nonlinear transformation $\bb{f}$ to the source distribution to obtain the target distribution, as $\bb{X}=\bb{f}(\bb{Z};\,\bb{W})$, where $\bb{f}$ is parameterised by a neural network with weights $\bb{W}$.
Invertibility allows us to compute the (log-)probability of any observed outcome 
$\bb{x}$ using the change-of-variables formula
\begin{align*}
\ln p_{\bb{X}}(\bb{x})
& = \ln p_{\bb{Z}}(\bb{f}^{-1}(\bb{x};\,\bb{W}))
+ \ln \vert \det \bb{J}_{\bb{f}^{-1}}(\bb{x};\,\bb{W}) \vert
\text{,}
\end{align*}
where $\bb{J}_{\bb{f}^{-1}}(\bb{x};\,\bb{W})$ is the Jacobian matrix of $\bb{f}^{-1}$ evaluated at $\bb{x}$.
Even if $\bb{f}$ is a relatively weak nonlinear transformation, we can obtain highly flexible distributions by chaining together several component transformations $\bb{f}_{l}$.
The combined transformation is often known as an \emph{invertible neural network}.
Crucially, both neural HMMs and normalising flows allow exact likelihood maximisation, making them a great fit for strong probabilistic models of speech acoustics.

\emph{Glow} \cite{kingma2018glow} is one popular normalising-flow architecture.
It interleaves learnt global affine transformations with so-called \emph{coupling layers}, that nonlinearly transform half of the elements of the input vector $\bb{z}_l\in\real^D$ with respect to the value of the remaining elements (which are left unaltered).
Calling the output of the $l$th coupling layer $\bb{z}_l^\prime$, the layer performs an element-wise invertible affine transformation that can be written
\begin{align*}
\bb{z}_{l,\,1:\nicefrac{D}{2}}^\prime
& = \bb{z}_{l,\,1:\nicefrac{D}{2}}\\
\bb{z}_{l,\,\nicefrac{D}{2}{+}1:D}^\prime
& = \bb{\alpha}_l(\bb{z}_{l,\,1:\nicefrac{D}{2}})
\odot \bb{z}_{l,\,\nicefrac{D}{2}{+}1:D}
+ \bb{\beta}_l(\bb{z}_{l,\,1:\nicefrac{D}{2}})
\text{,}
\end{align*}
where $\bb{\alpha}_l>\bb{0}$ and $\bb{\beta}_l$ are the outputs of a neural network with weights $\bb{W}$.
It is easy to verify that this transformation can be inverted, since $\bb{\alpha}_l$ and $\bb{\beta}_l$ can be computed by feeding $\bb{z}_{l,\,1:\nicefrac{D}{2}}^\prime$ into the neural net.
The matrix multiplication (a.k.a.\ ``1$\times$1 convolution'') in the affine transformation can be seen as a generalisation of a permutation operation \cite{kingma2018glow}, ensuring that all input elements receive multiple nonlinear transformations by the flow.

Most normalising flows in TTS are based on Glow, with Glow-TTS introducing a more parameter-efficient variant of the global affine transformations \cite{kim2020glow}.
%
Since both Glow and Glow-TTS use CNNs in
the coupling layers, the resulting invertible neural network has a finite receptive field.
As such, it cannot capture global dependencies between acoustics, which might explain the idiosyncratic utterance-level intonation and emphasis patterns generated by many trained Glow-TTS systems \cite{popov2021grad}.
In our proposal the source distribution $\bb{Z}_{t}$ depends on all previous outputs $\bb{z}_{1:t-1}$ through autoregression in a neural HMM (with long-range memory provided by an LSTM),
potentially producing more globally consistent synthetic speech.

\section{Experiments}
\label{sec:experiments}

To validate our proposal we compared three prior acoustic models to a comparable version of OverFlow in several experiments.
We also compared to an end-to-end and a non-parallel TTS baseline.
See \href{\webpageurl}{\webpageurltext} for audio.

The baseline acoustic models were Tacotron 2 \cite{shen2018natural} (label \textbf{T2}), Glow-TTS \cite{kim2020glow} (\textbf{GTTS}), and the neural HMM TTS (\textbf{NHMM}) system in \cite{mehta2022neural}.
For Tacotron 2, we used the Nvidia implementation\footnote{\href{https://github.com/NVIDIA/tacotron2/}{https://github.com/NVIDIA/tacotron2/}} while Glow-TTS used the official GitHub implementation\footnote{\href{https://github.com/jaywalnut310/glow-tts/}{https://github.com/jaywalnut310/glow-tts/}} and sampling temperature (the variance of $\bb{Z}_t$) 0.667 during synthesis.
Both used default hyperparameters, for which T2 and GTTS have a very similar model size.
The NHMM baseline used the same configuration as system ``NH2'' in \cite{mehta2022neural}, which inherits the encoder and decoder architectures from the T2 system (minus one of the decoder LSTMs).
NHMM used greedy deterministic output generation $\bb{x}_t=\bb{\mu}_t$, which is equivalent to random sampling with temperature zero.

For the OverFlow system in the experiments, we added an invertible network adapted from the official 
Glow-TTS code on top of the NHMM baseline architecture.
By reducing the number of nodes in each hidden layer in that added network from 192 to 150, we obtained a total model size very similar to the T2 and GTTS baselines; see Table \ref{tab:results}.
Our experiments opted for component networks with simple architectures that maximise comparability with prior system implementations, but it is worth noting that OverFlow is 
a framework within which many different network architectures can be used. 
It is, e.g., entirely possible to employ Transformers \cite{vaswani2017attention} inside any of the blocks in Fig.\ \ref{fig:overview}, or, for incremental TTS, one might swap in a causal encoder architecture with optional finite-horizon look-ahead.


All discrete-state acoustic models used two encoder vectors per phone, since this has been found to improve synthesis quality with both neural HMMs and Glow-TTS.
Tacotron 2, in contrast, does not need this, since it 
can represent arbitrary intermediate states using its continuous-valued attention instead.

\begin{figure}
\centering
\includegraphics[width=\columnwidth]{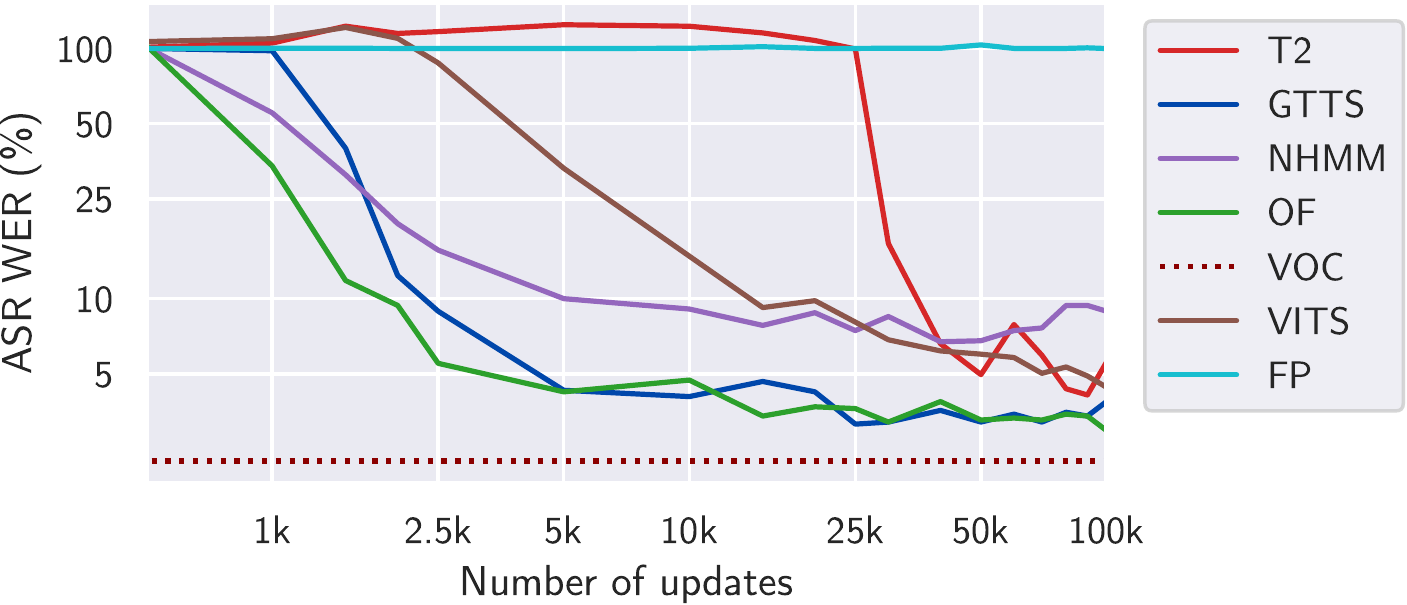}
\caption{Log-log plot of validation-set TTS WER over time.}
\label{fig:wer}
\vspace{-1.5\baselineskip}
\end{figure}

As additional baselines we included \textbf{VITS} \cite{kim2021vits}, an end-to-end model with normalising flows, and FastPitch v1.1 (\textbf{FP}) \cite{lancucki2021fastpitch,badlani2022one}, a widely used non-autoregressive neural TTS model.
Both of these used the implementation and hyperparameters from the Coqui TTS framework%
\footnote{\href{https://github.com/coqui-ai/TTS/}{https://github.com/coqui-ai/TTS/}}, with \texttt{espeak} for text processing.

We trained all systems on LJ Speech%
\footnote{\href{https://keithito.com/LJ-Speech-Dataset/}{https://keithito.com/LJ-Speech-Dataset/}} 
for 100k updates with batch size 32 on a single GPU.
Systems were provided the same normalised, phonetised text input and the same mel-spectrogram output features with the same vocoder for waveform generation, namely the pre-trained (V1) universal HiFi-GAN \cite{kong2020hifi} model%
\footnote{\href{https://github.com/jik876/hifi-gan/}{https://github.com/jik876/hifi-gan/}}
followed by a denoising filter as introduced in \cite{prenger2019waveglow} at a strength of 0.004.
Phone durations were generated deterministically and not sampled for any acoustic model; NHMM and OverFlow made use of quantile-based duration generation \cite{ronanki2016median,henter2017nonparametric}.
OverFlow is seen to improve modelling accuracy very substantially, since it reaches a validation-set per-sequence log-likelihood of 35k after training, compared to 6.5k for NHMM.

For the trained OverFlow system, we generated output 
under three \emph{conditions} that varied in sampling temperature and in the use of synthesis-time pre-net dropout.
The default condition (\textbf{OF}) uses a dropout of 0.5 (same as training) and a sampling temperature of 0.667; the no-dropout condition (\textbf{OFND}) set synthesis-time pre-net dropout to 0; whilst the zero-temperature condition (\textbf{OFZT}) instead set sampling temperature to 0, to allow comparing random sampling to greedy output generation.

Fig.\ \ref{fig:wer} graphs how the ASR word error rate (WER) of synthesising the 100 validation utterances evolves during training, as computed using Whisper
\cite{radford2022robust}.
WERs from modern ASR are known to correlate well with speech intelligibility to human listeners \cite{taylor2021confidence}.
As expected, T2 requires many more updates than other systems to learn to speak since it does not enforce monotonic alignments.
We also measured the wall-clock time until reaching 5\% WER on the validation utterances, and report the numbers in Table \ref{tab:results}.
Results show that Glow-TTS and OverFlow reached this level at least seven times faster than other systems.
Surprisingly, FastPitch did not learn to speak intelligibly at 100k, perhaps due to v1.1 not using an external aligner.

To evaluate intelligibility on phonetically balanced material, we synthesised the 720 Harvard sentences \cite{rothauser1969ieee}, giving the WERs listed in Table \ref{tab:results}.
The OverFlow conditions performed substantially better than all other baselines on these sentences.

\begin{table}[!t]
\centering
\begin{tabular}{@{}ll|cc|cccc@{}}
\toprule 
Me- & Con- & Temp- & Pre-net & Model & WER & Time & \tabularnewline
thod & dition & erature & dropout & size & {@}100k & to 5\% & MOS\tabularnewline
\midrule
\cite{shen2018natural} & T2 & N/A & \yes & 28.2M & 6.36\% & 54 h & 3.25\tabularnewline
\cite{kim2020glow} & GTTS & 0.667 & N/A & 28.6M & 3.97\% & \tablebf{2.5 h} & 2.64\tabularnewline
\cite{mehta2022neural} & NHMM & 0 & \yes & 15.3M & 5.96\% & 125 h & 2.97\tabularnewline
\cite{kim2021vits} & VITS & 0.667 & N/A & 83.1M & 7.03\% & 23 h & -\tabularnewline
\cite{lancucki2021fastpitch} & FP & N/A & N/A & 37.5M & 100\% & 21 h & -\tabularnewline
\midrule 
\multirow{3}{*}{\begin{turn}{90}\makecell{Prop.}\end{turn}} & OF & 0.667 & \yes & 28.5M & 2.91\% & 3 h & \tablebf{3.43}\tabularnewline
 & OFND & 0.667 &  & \textquotedbl & 2.92\% & \textquotedbl & 3.25\tabularnewline
 & OFZT & 0 & \yes & \textquotedbl & \tablebf{2.30\%} & \textquotedbl & 3.01\tabularnewline
\bottomrule
\end{tabular}
\caption{Summary of experiment results. 95\% confidence intervals for MOS values are $\pm$0.07. VOC MOS was 4.18$\pm$0.06.}
\label{tab:results}
\vspace{-2.5\baselineskip}
\end{table}

To evaluate subjective TTS quality, we carried out a web-based listening test on 40 sentences from the LJ Speech test set, enabling us to also compare to vocoded speech (condition \textbf{VOC}).
VITS and FP were excluded since VITS is not an acoustic model (and uses a different vocoder) whilst FP did not speak intelligibly at 100k updates.
We recruited 60 self-reported native speakers of English via \href{https://prolific.co/}{Prolific}, who all reported wearing headphones during the test.
The test used a MUSHRA-like \cite{itu2015method} design where listeners were presented with all seven versions of each utterance side by side on a single screen in random order and, similar to a MOS test \cite{itu1996telephone}, rated the naturalness of the audio samples on an integer scale from 1 (``Bad'') to 5 (``Excellent'').
Each participant rated 16 utterances, randomly selected from the pool of 40.
No test stimuli had any attention issues.
Stimuli were loudness-normalised to $-$20 dB LUFS based on EBU R128 \cite{ebu2020loudness}.
We additionally included two screens with attention checks, identical to the others except that one audio stimulus instructed the listener to rate that stimulus as 1.
Ratings from these screens were excluded from the analysis.
Listeners who failed the attention checks or rated VOC${\leq}2$ more than twice were disqualified and replaced with new listeners.


Mean opinion scores (MOS) from the listening test are found in Table \ref{tab:results}.
We see that our proposed addition of flows substantially improved on NHMM, and that OverFlow also performs much better than GTTS.
Pairwise Wilcoxon signed-rank tests find
all pairs of conditions in Table \ref{tab:results} except for (OFND, T2) to be
significantly different ($p<0.05$) after Holm-Bonferroni \cite{holm1979simple} correction.
OF thus performs better than T2, GTTS, and NHMM in the subjective evaluation, and in addition produces intelligible speech much sooner than T2, VITS, or FP, which is important during system development and tuning.

OverFlow is easy to fine-tune on diverse accents; see \href{\webpageurl}{our webpage} for examples.
It also achieves
good audio quality with both zero and nonzero sampling temperature
(OFZT vs.\ OF/OFND in our listening test).
This contrasts against Glow-TTS, where quality degrades substantially at temperature zero (see examples on \href{\webpageurl}{our webpage}), and against NHMM, which instead does much better if the temperature is zero (cf.\ \cite{mehta2022neural}).
Random sampling (OF) is also favoured over not sampling (OFZT), which indicates a highly accurate probabilistic model \cite{henter2014measuring}.


\section{Conclusions and future work}
\label{sec:conclusion}
We have described how the framework of neural HMMs in TTS can be combined with normalising flows (through an invertible post-net) to define strong probabilistic models of speech durations and acoustics.
The result is a neural transducer with normalising flows.
A simple example system based on our method rapidly learns to speak with fewer mispronunciations than comparable methods, and achieves strong subjective naturalness ratings.
%
Future work includes stronger models using Transformers \cite{vaswani2017attention,li2019neural},
multi-speaker synthesis, and applications to diverse and challenging data such as spontaneous speech \cite{szekely2019spontaneous,lameris2023prosody}.

\ifinterspeechfinal
\section{Acknowledgements}
\label{sec:acks}
This work was partially supported by the Wallenberg AI, Autonomous Systems and Software Program (WASP) funded by the Knut and Alice Wallenberg Foundation, the Swedish Research Council projects
VR-2019-05003
and
VR-2020-02396
and Riksbankens Jubileumsfond project P20-0298.
\else
\fi

\bibliographystyle{IEEEtran}
\bibliography{refs_abbrev}

\end{document}